\documentclass[twocolumn,floatfix,prb,showpacs]{revtex4}
\usepackage{graphicx,graphics,color,epsfig}
\usepackage{bm}
\usepackage{amsmath}
\usepackage{amssymb}

\begin{document}

\title{Contractor renormalization group theory of the SU($N$) chains and
ladders}
\author{Peng Li and Shun-Qing Shen}
\affiliation{Department of Physics, The University of Hong Kong, Pokfulam Road, Hong
Kong, China}
\date{\today}

\begin{abstract}
Contractor renormalization group (CORE) method is applied to the SU($N$)
chain and ladders in this paper. In our designed schemes, we show that these
two classes of systems can return to their original form of Hamiltonian
after CORE transformation. Successive iteration of the transformation leads
to a fixed point so that the ground state energy and the energy gap to the
ground state can be deduced. The result of SU($N$) chain is compared with
the one by Bethe ansatz method. The transformation on spin-1/2 ladders gives
a finite gap in the excited energy spectra to the ground state in an
intuitive way. The application to SU(3) ladders is also discussed.
\end{abstract}

\maketitle

The contractor renormalization (CORE) group method combines the contraction
and cluster expansion techniques with the real space renormalization group
approach to solve the electron and spin lattice problems.\cite{MW1} It was
first applied to spin-1/2 Heisenberg chain and (1+1)-dimensional Ising
model, and later to the frustrated antiferromagnets and the Haldane
conjecture. The results are satisfactory and encouraging.\cite{MW1,W1} Since
then the method has been applied to investigate low energy physics in many
strongly correlated systems. \cite{AA1,CP1,BAA1,PSC1,CLM1,BA1} In this
paper, we are concerned with a class of models showing that CORE is at its
\emph{critical point}, which means that the symmetry of the system is
restored or the same form of Hamiltonian is reconstructed after the CORE
transformation, just like the spin-1/2 Heisenberg chain.\cite{MW1}
Undoubtedly, this method is not limited to such a kind of systems. Though in
many systems the original Hamiltonian cannot be recovered, the lower energy
physics are retained and studied successively after the truncation and
transformation.

We consider the SU($N$) chain and ladders in this work. We show that the
same form of the Hamiltonian is recovered after dividing adequately the
lattice into blocks and defining a truncation scheme, so that the CORE
algorithm can be done recursively. The range-2 result for SU($N$) chain is
compared with the Bethe ansatz solution by Sutherland.\cite{S1} The
comparison suggests that the CORE method can give good result especially for
large $N$ and relatively larger blocks. We also present the results of the
real-space renormalization group (RG) theory, which usually agrees
qualitatively with the one by range-2 CORE calculation.\cite{W1} In many
cases the latter can be regarded as a refined method on the former. The
spin-1/2 ladders have attracted a lot of attention since the discovery of a
finite spin gap in the 2-leg ladders.\cite{DRS1,DR1} Another CORE scheme
based on plaquette-dividing of the ladder had been applied to this system.%
\cite{PS1,CLM1} Here we shall use a different scheme which shows a $S=1$
magnon gap in an intuitive way. Results up to range-3 are presented.

It was shown by Morningstar and Weinstein that the CORE scheme of three-site
block partition and two-state truncation on SU(2) chain recover the original
form of Hamiltonian. Then the resulting effective Hamiltonian can be solved
iteratively and a quite satisfactory result can be obtained.\cite{MW1} The
recovery of the form of the Hamiltonian owes highly to the SU(2) symmetry
and an adequate designed CORE scheme. As a generalization, we found that
their CORE scheme on SU(2) chain is a speciman picked out from a general
CORE scheme on the SU($N$) chain. Though the SU($N$) chain had been exactly
solved by Bethe ansatz method long time ago,\cite{S1} it is still
instructive to see how CORE works in the system.

Let us start with a one-dimensional SU($N$) chain in terms of the exchange
operator, $H=J\sum_{j}P_{j,j+1}$. Here we limit our discussion to the
antiferromagnetic case by setting $J=1>0$. For a SU($N$) system each site $j$
has $N$ quantum states $|j,\alpha \rangle $ with $(\alpha =1,2,\cdots ,N)$.
The exchange operator $P_{j,j+1}$ swaps two states on sites $j$ and $j+1$,
i.e. $P_{j,j+1}|j,\alpha ;j+1,\beta \rangle =|j,\beta ;j+1,\alpha \rangle $.
Usually $P_{j,j+1}$ can be expressed in terms of the SU($N$) generators as $%
P_{j,j+1}=\sum_{\alpha \beta }J_{\beta }^{\alpha }(j)J_{\alpha }^{\beta
}(j+1),$ where the operators $J_{\beta }^{\alpha }(j)$ satisfy the SU($N$)
algebra $\left[ J_{\beta }^{\alpha }(j),J_{\nu }^{\mu }(j\prime )\right]
=\delta _{jj\prime }\left( \delta _{\nu }^{\alpha }J_{\beta }^{\mu
}(j)-\delta _{\beta }^{\mu }J_{\nu }^{\alpha }(j)\right) $. Alternatively, $%
P_{j,j+1}$ can also be expressed by spin operators.\cite{Schrodinger1,Li04}
Many spin systems as well as spin-obital systems concerning SU($N$) symmetry
have been studied extensively.\cite{AA2,RS1,LMSZ1,S,PMFM1}

In the CORE scheme, the first step is to divide the original chain into a
chain of blocks and retain adequate number of energy levels in each block.
We found two obvious schemes to be readily applied to this system: one is $%
(N-1)$-site block patition with $N$-state truncation (scheme A), and the
other is $(N+1)$-site block patition with $N$-state truncation (scheme B).
The treatment on SU(2) case in Ref. \cite{MW1} obviously falls into the
scheme B with $P_{j,j+1}=2\mathbf{S}_{j}\cdot \mathbf{S}_{j+1}+1/2$ when $%
N=2 $. We will see the scheme B gives better results than the scheme A. The
existence of the two schemes can be understood from the single column Young
tableaux with $(N-1)$ or $(N+1)$ boxes. In fact the SU($N$) model on both $%
(N-1)$-site block and $(N+1)$-site block have one unique $N$-dimensional
ground state space. We denote the truncated space for a single block by $%
\Phi _{j}=\{\left\vert \phi _{j,1}\right\rangle ,\left\vert \phi
_{j,2}\right\rangle ,\cdots ,\left\vert \phi _{j,N}\right\rangle \}$. Then
in the range-2 CORE calculation, we should retain appropriate $N^{2}$ low
levels from the exact diagonalization of two blocks. All the retained low
levels\ should have nonzero projection to the product space $\Phi
_{j}\otimes \Phi _{j+1}$, so the eligible levels are not always the lowest
ones. Fortunately this job is easy to be done due to the SU($N$) symmetry.
The range-2 CORE calculation leads to the effective Hamiltonian,%
\begin{equation}
H^{(2)}=\frac{1}{N\mp 1}\sum\limits_{j}(-C_{\mp }+K_{\mp }\widetilde{P}%
_{j,j+1}),  \label{Hr2}
\end{equation}%
where the sign $\mp $ corresponds to the two schemes A ($-$) and B ($+$), $%
\widetilde{P}_{j,j+1}$ is a "new" renormalized exchange operator connecting
blocks $j$ and $j+1$ after each block "contracts" to a single site. The
coefficients $C_{\mp }$ and $K_{\mp }$ are listed in Table \ref{tCK}. It can
be confirmed the range-3 Hamiltonian will include another operator $%
\widetilde{P}_{j,j+2}$ and the range-4 Hamiltonian will include more
operators like $\widetilde{P}_{j,j+3},\ \widetilde{P}_{j,j+1}\widetilde{P}%
_{j+2,j+3},\ \widetilde{P}_{j,j+2}\widetilde{P}_{j+1,j+3},\ \widetilde{P}%
_{j,j+3}\widetilde{P}_{j+1,j+2}$. Here we only give the range-2 results
since higher range calculation will not change the physics. For the scheme A
and B, we give the results for $N=3,4,5$ and $2,3,4$ respectively.

\begin{table}[tbp]
\begin{tabular}{l|cc|cc}
\hline
Scheme A & \multicolumn{2}{|l|}{RG} & \multicolumn{2}{|l}{Range-2 CORE} \\
\multicolumn{1}{c|}{$N$} & $C_{-}$ & $K_{-}$ & $C_{-}$ & $K_{-}$ \\ \hline
\multicolumn{1}{c|}{$3$} & $\frac{3}{4}$ & $\frac{1}{4}$ & $1.0731$ & $%
0.3411 $ \\
\multicolumn{1}{c|}{$4$} & $\frac{16}{9}$ & $\frac{1}{9}$ & $2.2485$ & $%
0.1787$ \\
\multicolumn{1}{c|}{$5$} & $\frac{45}{16}$ & $\frac{1}{16}$ & $3.3762$ & $%
0.1112$ \\ \hline\hline
Scheme B & \multicolumn{2}{|l|}{RG} & \multicolumn{2}{|l}{Range-2 CORE} \\
\multicolumn{1}{c|}{$N$} & $C_{+}$ & $K_{+}$ & $C_{+}$ & $K_{+}$ \\ \hline
\multicolumn{1}{c|}{$2$} & $\frac{13}{18}$ & $\frac{4}{9}$ & $0.9956$ & $%
0.4916$ \\
\multicolumn{1}{c|}{$3$} & $2.1693$ & $0.2654$ & $2.5982$ & $0.3084$ \\
\multicolumn{1}{c|}{$4$} & $3.4111$ & $0.1728$ & $3.9432$ & $0.2089$ \\
\hline
\end{tabular}%
\caption{The coefficients $C_{\mp }$ and $K_{\mp }$ in Eq. (\protect\ref{Hr2}%
)}
\label{tCK}
\end{table}

Successive application of CORE on Eq. (\ref{Hr2}) will lead the running
coupling approaching a gapless fixed point. And no phase transition is
observed. The ground energy is read out as%
\begin{equation}
E_{0}=-\frac{C_{\mp }}{(N\mp 1)-K_{\mp }}.  \label{E0}
\end{equation}%
where the sign $\mp $ corresponds to the two schemes. Fig. 1 shows that the
result of the range-2 CORE of the scheme B agrees quite well with the one by
Bethe ansatz method. The numerical error can be reduced by higher range
calculation. The range-4 result for $N=2$ by Weistein shows the error is
reduced to $-0.0025$.\cite{W1}

\begin{figure}[tbp]
\includegraphics[width=7cm]{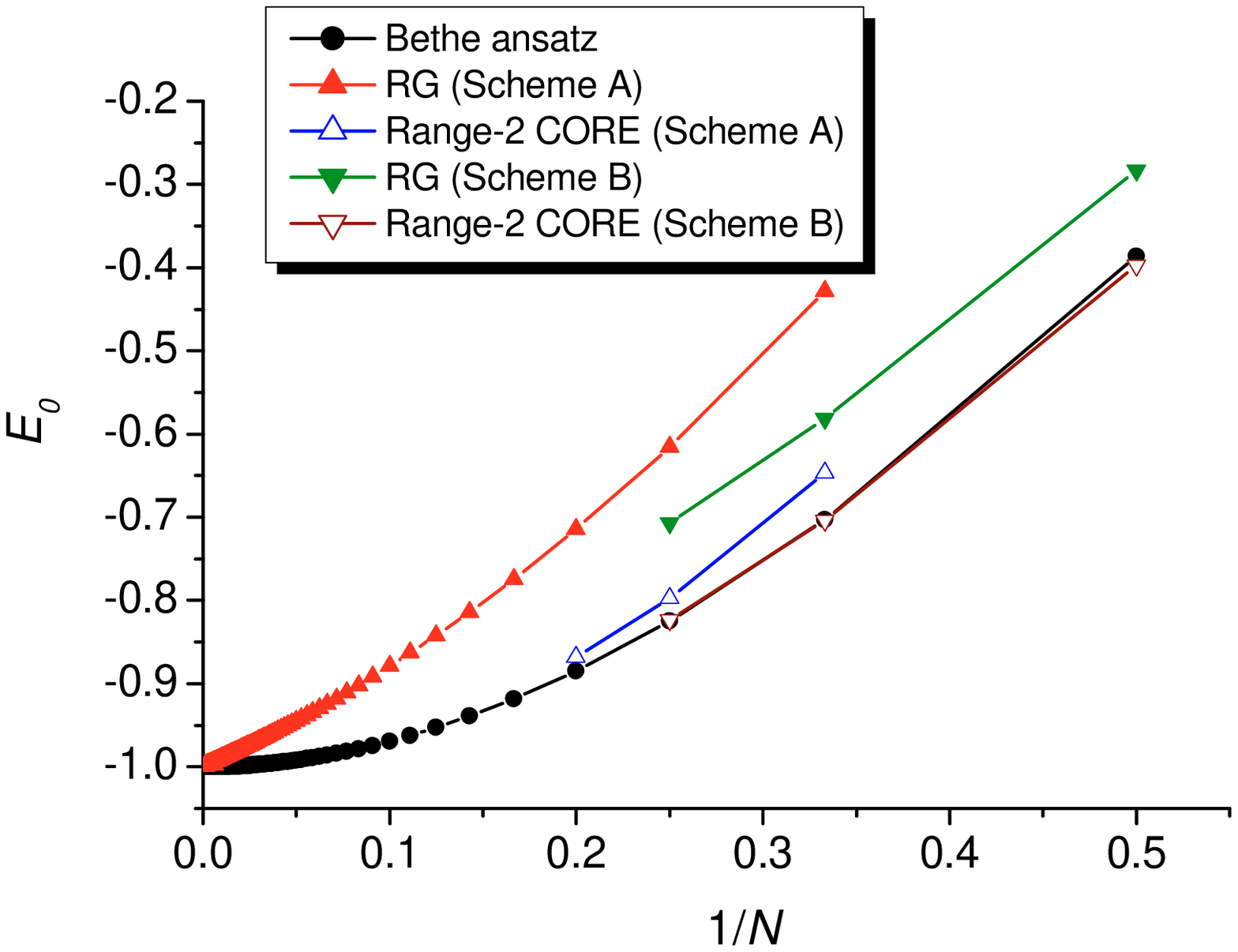}\newline
\caption{(Color online) The ground energy of SU($N$) chain. The
scheme B of CORE gives better results and the numerical errors
are about $-0.0106,-0.0006,0.0021$ for $N=2,3,4$ respectively
compared to the results by Bethe ansatz \cite{S1}.} \label{FIG1}
\end{figure}

In fact the traditional RG gives an effecive Hamiltonian having the same
form of Eq. (\ref{Hr2}). It can produce results consistent with CORE though
not so good.\cite{W1} The two schemes above are still applicable and the
corresponding coefficients can be found in Table \ref{tCK}. The advantages
of CORE are obvious. In many cases one can design more flexible schemes in
CORE while selecting basic blocks and truncating at low levels.\cite%
{AA1,CP1,BAA1,PSC1,CLM1,BA1} A more careful analysis shows that RG based on (%
$N-1$)-site block partition scheme (scheme A) can give an effective
Hamiltonian for general $N$,%
\begin{equation}
H^{RG}=\frac{1}{N-1}\sum\limits_{j}\left[ -\frac{N(N-2)^{2}}{(N-1)^{2}}+%
\frac{1}{(N-1)^{2}}\widetilde{P}_{j,j+1}\right] ,  \label{Hnr}
\end{equation}%
which exhibits a ground energy coinciding with the one by Bethe ansatz
method at large $N$, $E_{0}=-N(N-2)/(N^{2}-N+1)\overset{N\rightarrow \infty }%
{\longrightarrow }-1$.

The 2-leg spin-1/2 ladders aroussed a lot of attention when a finite spin
gap was observed.\cite{DR1} A simple picture says that the ground state is a
product state with the spins on each rung forming a spin singlet. Then the
lowest energy excitation is a $S=1$ magnon. Here we show that our scheme of
CORE produces exactly the same picture and refined results can be achieved
following the CORE algorithm. We start from the Hamiltonian,%
\begin{equation}
H=\sum_{j}\left[(\mathbf{S}_{j}^{A}\cdot \mathbf{S}_{j+1}^{A}+\mathbf{S}%
_{j}^{B}\cdot \mathbf{S}_{j+1}^{B})+ \alpha \ \mathbf{S}_{j}^{A}\cdot
\mathbf{S}_{j}^{B}\right] ,  \label{Hl}
\end{equation}%
where the indices $A$ and $B$ refer to the two rails of the ladders, $\alpha
=J_{\text{rung}}/J_{\text{rail}}$ is the ratio between the rung and rails
couplings, and we have set $J_{\text{rail}}=1$.

\begin{figure}[tbp]
\includegraphics[width=6cm]{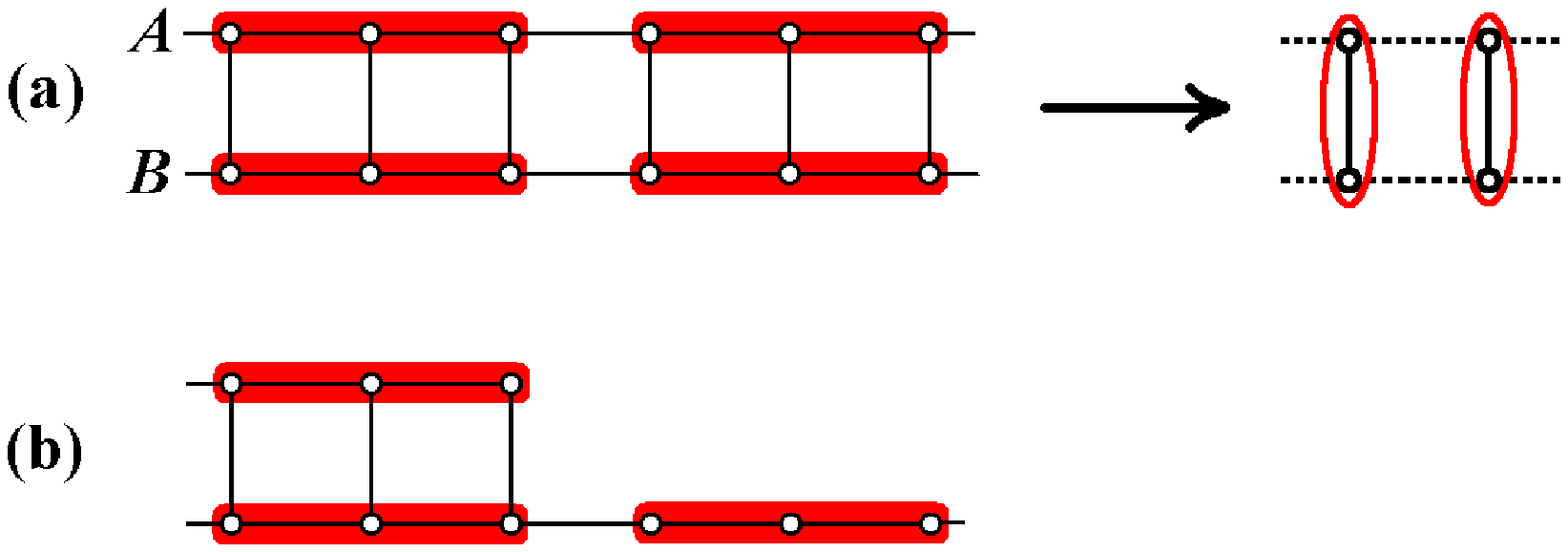}\newline
\caption{(Color online) (a) 2-leg ladders. The basic blocks are triads along
the rail direction. The fixed point is a chain of decoupled dimers. (b) The
unsymmetric configuration of blocks involved in range-3 CORE calculation.}
\label{FIG2}
\end{figure}

Our first step is to divide the ladder into triads along the rail direction
(Fig. 2(a)). The problem on the rail direction is just the SU(2) chain that
had been solved. Detailed calculation shows that the effective interaction
between the two blocks along the rung can also recover the Heisenberg
interaction. Thus "new" ladders with renormalized couplings can be obtained.
The second step is to parse out the effective block-block interactions from
all possible configurations of connected blocks.\ As defined by Morningstar
and Weinstein, $r$ connected blocks contain range-$r^{\prime }$ interactions
with $r^{\prime }=0,1,\cdots ,r$ ($r^{\prime }=0$ corresponds to the
constant term as in Eq. (\ref{Hlr2})).\ To parse out all range interactions
the exact diagonalization is employed on the connected blocks. We present
range-2 and range-3 results here. It is notable that the range-3 blocks
should include a configuration in Fig. 2(b). This unsymmetric configuration
may make the iteration procedure more troublesome.

The range-2 CORE result simply regains the original form of Hamiltonian
except for a constant term,
\begin{equation}
H^{(2)}=\frac{1}{3}\sum_{j}[-C(\alpha )+\delta \ (\widetilde{\mathbf{S}}%
_{j}^{A}\cdot \widetilde{\mathbf{S}}_{j+1}^{A}+\widetilde{\mathbf{S}}%
_{j}^{B}\cdot \widetilde{\mathbf{S}}_{j+1}^{B})+\Lambda (\alpha )\
\widetilde{\mathbf{S}}_{j}^{A}\cdot \widetilde{\mathbf{S}}_{j}^{B}],
\label{Hlr2}
\end{equation}%
where $\delta =0.491582$, $C(\alpha )$ and $\Lambda (\alpha )$ vary with $%
\alpha $. The iteration on the range-2 effective Hamiltonian is always
applicable because the retained four low levels are always one spin singlet
and three spin triplets just like the SU(2) chain case. After $n$ steps of
iteration on Eq. (\ref{Hlr2}) we will get running coupling terms as $%
h_{n}=\delta _{n}(\widetilde{\mathbf{S}}_{j}^{A}\cdot \widetilde{\mathbf{S}}%
_{j+1}^{A}+\widetilde{\mathbf{S}}_{j}^{B}\cdot \widetilde{\mathbf{S}}%
_{j+1}^{B})+\Lambda _{n}(\alpha )\ \widetilde{\mathbf{S}}_{j}^{A}\cdot
\widetilde{\mathbf{S}}_{j}^{B}$, where the coefficients are determined
recursively, $\delta _{n}=\delta ^{n},\Lambda _{n}(\alpha )=\delta ^{n-1}\
\Lambda (\frac{\Lambda _{n-1}(\alpha )}{\delta ^{n-1}}),\cdots \cdots
,\Lambda _{2}(\alpha )=\delta \ \Lambda (\frac{\Lambda (\alpha )}{\delta }%
),\Lambda _{1}(\alpha )=\Lambda (\alpha ),\Lambda _{0}(\alpha )=\alpha .$ So
the rail coupling approaches zero $\delta ^{n}\rightarrow 0$ as $%
n\rightarrow \infty $, while the rung coupling goes to a fixed value $%
\Lambda _{n\rightarrow \infty }(\alpha )\neq 0$ for $\alpha >0$ (we observed
that $\Lambda _{n\rightarrow \infty }(\alpha )\rightarrow 0$ only when $%
\alpha =0$, which is in agreement with the conclusion drawn by DMRG \cite%
{DRS1,LFS} and exact diagonalization \cite{Wa}). So the system flows to a
fixed point exhibiting dimer covering on each rung of the ladder. The spin
gap is read out as $\Delta _{s}(\alpha )=\Lambda _{n\rightarrow \infty
}(\alpha )$. The ground energy $E_{0}$ is obtained by cumulating the
constant term. Table \ref{tE} gives an example of iterations procedure for $%
\alpha =1$.

\begin{table}[tbp]
\begin{tabular}{cccc}
\hline
$n$ & $E_{0}$ & $\delta _{n}$ & $\Lambda _{n}(\alpha )$ \\ \hline
$0$ & \multicolumn{1}{l}{$0$} & \multicolumn{1}{l}{$1.0$} &
\multicolumn{1}{l}{$1.0$} \\
$1$ & \multicolumn{1}{l}{$-0.460796$} & \multicolumn{1}{l}{$0.491582$} &
\multicolumn{1}{l}{$0.81919$} \\
$2$ & \multicolumn{1}{l}{$-0.558041$} & \multicolumn{1}{l}{$0.241653$} &
\multicolumn{1}{l}{$0.64499$} \\
$5$ & \multicolumn{1}{l}{$-0.587214$} & \multicolumn{1}{l}{$0.028706$} &
\multicolumn{1}{l}{$0.420144$} \\
$10$ & \multicolumn{1}{l}{$-0.587867$} & \multicolumn{1}{l}{$0.000824$} &
\multicolumn{1}{l}{$0.382715$} \\
$15$ & \multicolumn{1}{l}{$-0.587869$} & \multicolumn{1}{l}{$0.000024$} &
\multicolumn{1}{l}{$0.381603$} \\
$20$ & \multicolumn{1}{l}{$-0.587869$} & \multicolumn{1}{l}{$6.7907\times
10^{-7}$} & \multicolumn{1}{l}{$0.381571$} \\
$21$ & \multicolumn{1}{l}{$-0.587869$} & \multicolumn{1}{l}{$3.3382\times
10^{-7}$} & \multicolumn{1}{l}{$0.38157$} \\
$22$ & \multicolumn{1}{l}{$-0.587869$} & \multicolumn{1}{l}{$1.641\times
10^{-7}$} & \multicolumn{1}{l}{$0.38157$} \\ \hline
\end{tabular}%
\caption{An example of the range-2 CORE iteration procedure at $\protect%
\alpha =1$.}
\label{tE}
\end{table}

The range-3 CORE result at the first run of iteration contains the
next-nearest-neighbour interactions,%
\begin{gather}
H^{(3)}=\frac{1}{3}\sum_{j}[-C(\alpha )+\delta (\alpha )\ (\widetilde{%
\mathbf{S}}_{j}^{A}\cdot \widetilde{\mathbf{S}}_{j+1}^{A}+\widetilde{\mathbf{%
S}}_{j}^{B}\cdot \widetilde{\mathbf{S}}_{j+1}^{B})  \notag \\
+\Lambda (\alpha )\ \widetilde{\mathbf{S}}_{j}^{A}\cdot \widetilde{\mathbf{S}%
}_{j}^{B}+\Omega (\alpha )\ (\widetilde{\mathbf{S}}_{j}^{A}\cdot \widetilde{%
\mathbf{S}}_{j+1}^{B}+\widetilde{\mathbf{S}}_{j}^{B}\cdot \widetilde{\mathbf{%
S}}_{j+1}^{A})  \notag \\
+\gamma \ (\widetilde{\mathbf{S}}_{j}^{A}\cdot \widetilde{\mathbf{S}}%
_{j+2}^{A}+\widetilde{\mathbf{S}}_{j}^{B}\cdot \widetilde{\mathbf{S}}%
_{j+2}^{B})],
\end{gather}%
where $C(\alpha ),\delta (\alpha )$ and $\Lambda (\alpha )$ are different
from the ones in Eq. (\ref{Hlr2}), $\gamma =0.033975$. $\gamma $ will vary
with $\alpha $ in the successive iterations , $\gamma _{n}=\gamma
_{n}(\alpha ),\cdots ,\gamma _{1}=\gamma $. After $n$-step iterations, we
found that the only nonvanishing coupling is still the interaction along the
rung $\Lambda _{n\rightarrow \infty }(\alpha )\neq 0$, so the physical
picture obtained by the range-2 CORE does not change, i.e. the ground energy
and the spin gap are produced in the same way.

\begin{figure}[tbp]
\includegraphics[width=8cm]{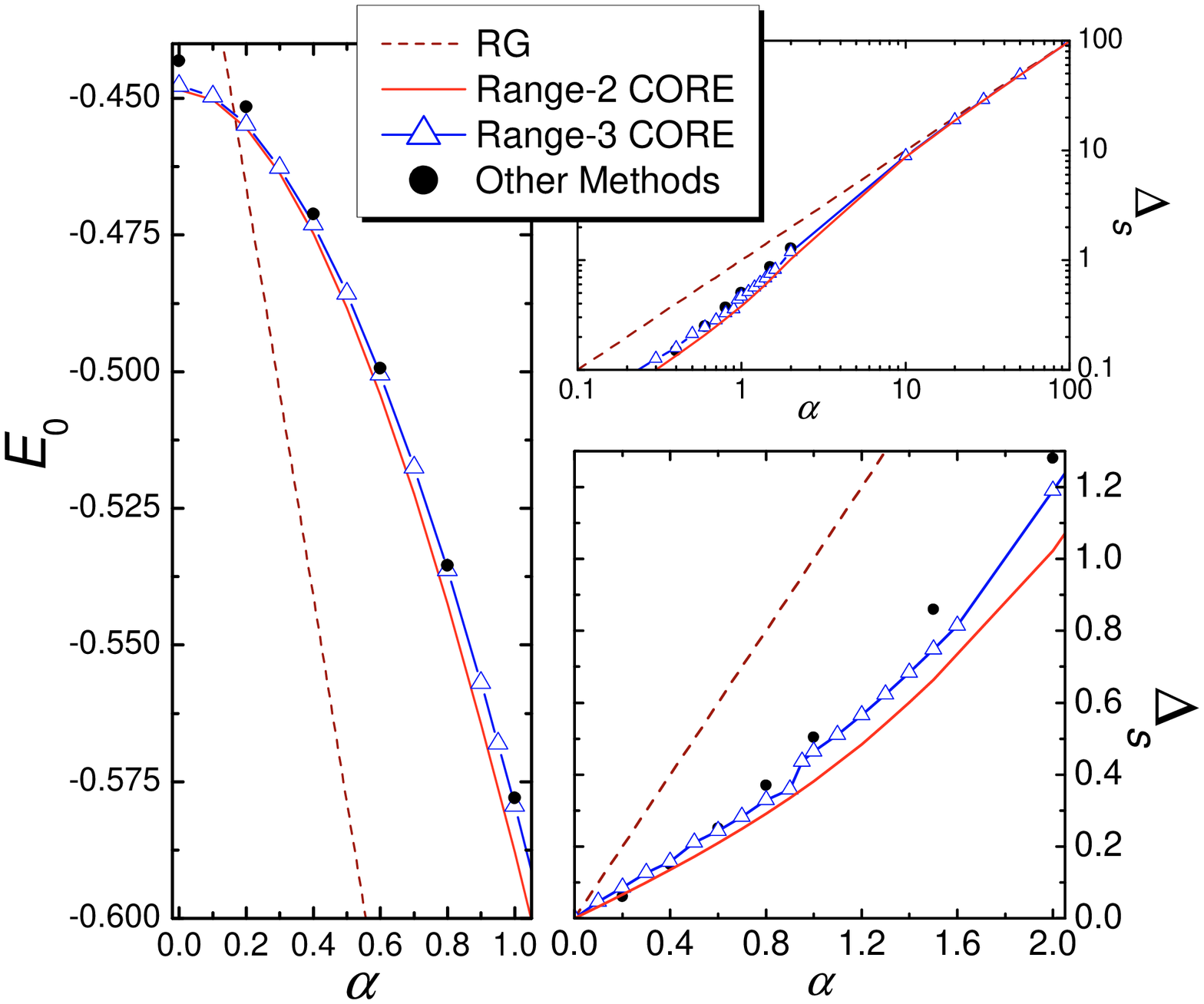}\newline
\caption{(Color online) The ground energy and the gap for the
spin-1/2 2-leg ladder. The log-log plot shows that CORE and RG
give correct gap in strong coupling limit $\protect\alpha
\rightarrow
\infty $. Data by other methods are adapted from Ref. \protect\cite%
{DR1,BDRS,WNS,F}.}
\label{FIG3}
\end{figure}

As we noted above, the unsymmetric configuration of blocks in Fig. 2(b)
brings some troubles to the range-3 CORE iteration. Unlike the SU(2) chain,
the desired low levels may not always stay at the lowest positions during
the iterations. And sometimes it is hard to select out the eligible set of
levels from several possible candidates since each of them will lead to a
recovered SU(2) symmetry. So different iteration procedures with different
results are inevitable. When these situations take place, we resort to the
principle: \emph{retaining the iteration procedure that gives the lowest
energy},\cite{WLS} although in our observations the values of the results
only have small difference. The range-2 and range-3 CORE results for the
ground state energy and the spin gap are illustrated in Fig. 3. For a
comparison, data by other methods \cite{DR1,BDRS,WNS,F} are presented
together. The ground energy agrees well with those by other methods in the
whole range of interchain coupling $\alpha $. This means that CORE algrithm
can successively capture the low energy physics of the system. The gap has
relatively larger deviation at intermediate values of $\alpha $.
Nevertheless the discrepancy can be remedied through higher range CORE
calculation. The range-3 gap is a little zigzag. This may be due to the
unsymmetric configuration of range-3 blocks in Fig. 2(b). It is noteworthy
that RG gives a gap simply as $\Delta _{s}^{RG}=\alpha $, which captures the
correct behaviour of the gap at strong coupling limit $\alpha \rightarrow
\infty .$\cite{DR1}

\begin{figure}[tbp]
\includegraphics[width=6.5cm]{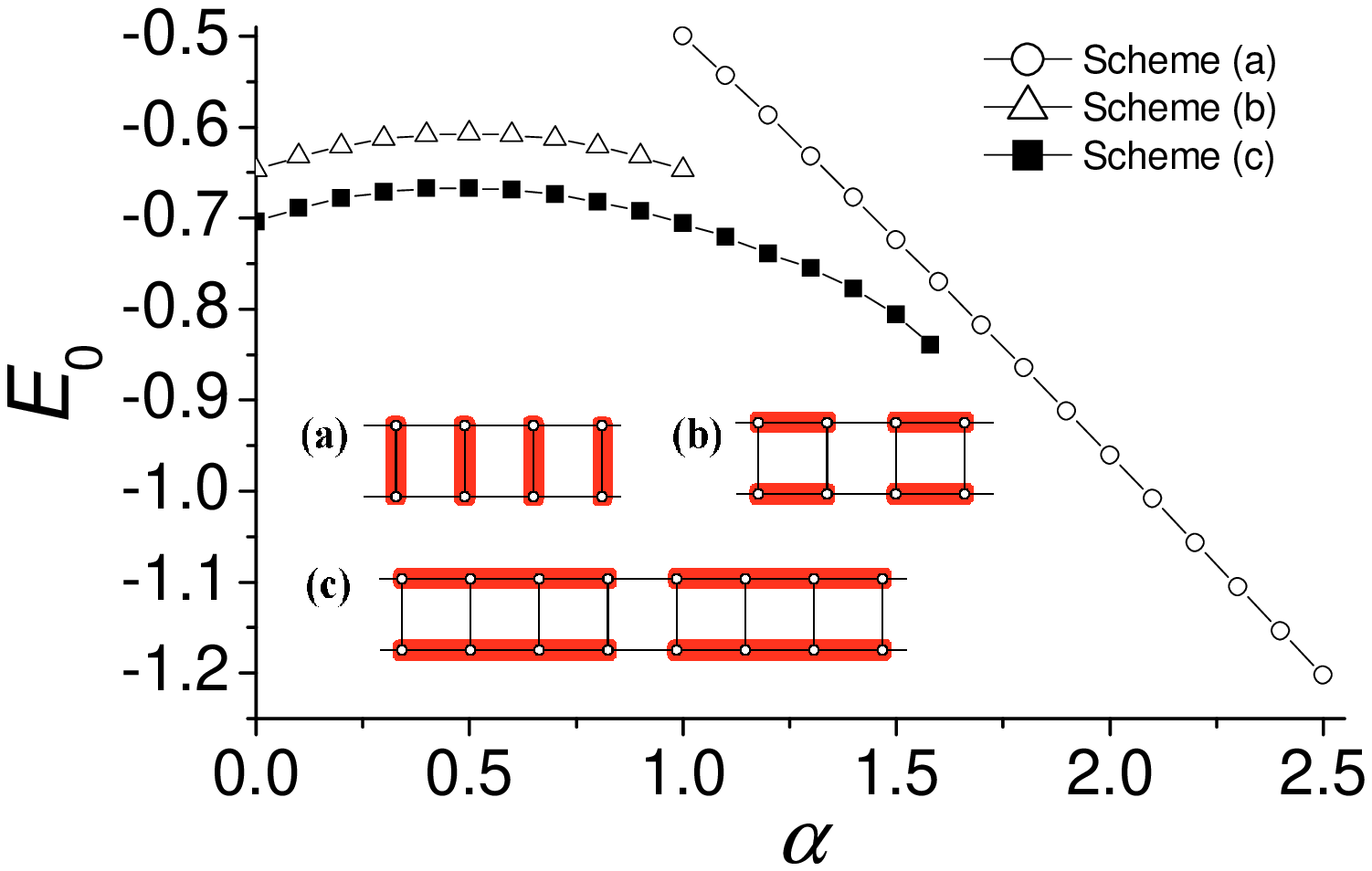}\newline
\caption{(Color online) The ground energy of the 2-leg SU(3)
ladders. The insets (a),(b) and (c) show the schemes used in the
calculation .} \label{FIG4}
\end{figure}

We also applied CORE to the 2-leg SU(3) ladders, $H=\sum_{j}\left[
(P_{j,A;j+1,A}+P_{j,B;j+1,B})+\alpha \ P_{j,A;j,B}\right] $. The applicable
schemes are presented in Fig. 4(a), (b) and (c). Notice that all blocks are
equivalent and a 3-state truncation is made in each scheme. The scheme (a)
should be valid when the rung interaction $\alpha $ is large enough. While
for small $\alpha $, the schemes (a) ($\alpha <1.0$) and (b) ($\alpha <1.58$%
) are appropriate and scheme (b) is better than (a). All the
three schemes lead to the fixed point with zero gap. We see that
scheme (a) will be mapped to a SU(3) chain, which had been solved
previously and gives a zero gap. And after the first mapping we
applied 4-site block partition scheme on the chain in the
successive iteration steps to produce the ground energy in Fig.
4. While scheme (b) and (c) will return to a new 2-leg SU(3) ladders, $%
h_{n}=\delta _{n}(P_{j,A;j+1,A}+P_{j,B;j+1,B})+\Lambda _{n}(\alpha )\
P_{j,A;j,B}$, but we observed that the running couplings of the rail
direction $\delta _{n}$ and the rung direction $\Lambda _{n}$ will go to
infinitesimals of the same order \cite{EXPLAIN} as we push the iteration
steps to infinity, $n\rightarrow \infty $, so a gapless phase is also
obtained, which agrees with the result of scheme (a). The SU(3) model on
4-leg ladders can be analyzed in similar schemes and a gapless result is
also expected. The result is reminiscent of the SU(2) model on a chain and
on a 3-leg ladders, which are also gapless. But unfortunately the above
schemes or their analogues are not applicable for the 2-leg SU(4) ladders,
which exhibites plaquette singlet-multiplet excitation.\cite{S,BDRS,BALM}
One may have to resort to other kind of schemes.

In conclusion, we have studied the SU($N$) chain and ladders by the CORE
schemes. We have shown that the effective Hamiltonian in the appropriate
CORE schemes can regain its original form such that it approaches a fixed
point by iteration of the CORE\ schemes. The ground state energy and the
lowest excitations can be deduced from the fixed point. The results show
that the SU($N$) chain and the 2-leg SU(3) ladders are gapless, while the
2-leg spin-1/2 ladder exhibites gapped phase originated from the rung
dimmerization.

This work was supported by the Research Grant Council of Hong Kong under the
project No. HKU7038/04P.


\begin{thebibliography}{99}
\bibitem{EXPLAIN} In fact the ratio $\Lambda _{n}/\delta _{n}$ will reach at
a fixed value $0.3597$ for scheme (b) and $1.1176$ for scheme (c) as $%
n\rightarrow \infty $, although $\Lambda _{n}\rightarrow 0$ and $\delta
_{n}\rightarrow 0$.
\end{thebibliography}
\end{document}